Letter

South Korean degenerative spondylolisthesis patients had surgical treatment at earlier age than Japanese, American, and European patients: a published literature observation

Running title: degenerative spondylolisthesis surgical treatment

Zoltan Kaplar[1], Yi-Xiang J Wang[1]

[1] Department of Imaging and Interventional Radiology, The Chinese University of Hong Kong, Shatin, New Territories, Hong Kong SAR, China.

Correspondence to: Yi-Xiang Wang, PhD, MMed, Dipl-Rad, Department of Imaging and Interventional Radiology, The Chinese University of Hong Kong, Shatin, New Territories, Hong Kong SAR
E-mail: yixiang_wang@cuhk.edu.hk

## Abstract

Physical therapy is the first line of treatment for adults with symptoms from degenerative spondylolisthesis. Surgical management is offered when nonoperative options have not adequately relieved symptoms. We performed PubMed literature search with the word 'spondylolisthesis', and updated till September 18, 2016. We selected original research data involving surgical treatment of degenerative spondylolisthesis, and in total there were n articles, including data reported from Japan (n=37 series), South Korea (n=11 series), mainland China (n=5 series), and ROC (Taiwan, n= 3 series), America (n=20 series), Europe (n=23 series). The mean age of each study were extracted and used a single entry. We tried our best to filter out double/multiple reported data. Our results showed the median age of degenerative spondylolisthesis patients underwent surgical treatment was 66 years in Japan, 60 years in South Korea, 59 years in mainland China, 59 years in ROC (Taiwan), 65 years in USA, and 66 years in Europe. This study indicates Japan and South Korea may have different surgical practice pattern. It will be of interested to investigate whether more proportion of degenerative spondylolisthesis patients have been treated surgically in South Korea than in Japan. The cost-effectiveness of different approaches may warrant further analysis by professional spine surgeons.

The epidemiology of lumbar degenerative spondylolisthesis (DS) remains controversial. We recently performed a systemic review with the aim to have a better understanding of DS's prevalence in general population. The results showed the prevalence of DS is very gender specific and age specific [1]. Both women and men have few DS before 50 years old, after 50 years old both women and men start to develop DS, with women having a faster developing rate than men. For elderly Chinese (mean age: 72.5 yrs), DS prevalence is around 25.0% for women and 19.1% for men, and the prevalence women:men ratio is 1.3:1. Elderly Caucasian American may have a higher DS prevalence, being approximately 60-70% higher than elderly Chinese; however the prevalence women:men ratio was similar to elderly Chinese population [1].

The majority of symptomatic degenerative spondylolisthesis patients are successfully treated without surgery. Physical therapy is the first line of treatment for adults with symptoms from spondylolisthesis. Hamstring stretching, trunk strengthening, and avoidance of inciting activities are beneficial for adults. Surgical management is offered when nonoperative options have not adequately relieved symptoms. Patients for whom surgery is indicated usually have good outcomes. Young patients may require only a fusion *in situ*; however, patients who have evidence of neural compression may need a decompression to relieve symptoms, and fusion is usually also indicated in these cases [2, 3].

In our last study [1], preliminary data showed the ratio of numbers of female patients received treatment compared with men did not differ between Northeast Asians (Chinese, Japanese, and Korean) and European and American Caucasians, being around 2:1 in elderly population. However, compared with Caucasians, Asians were likely to have surgical treatment more than half decade earlier [1]. We were interested whether it was due to Northeast Asians manifest clinical symptoms earlier or more severe, and therefore did this further literature analysis. We used the PubMed search results we obtained for our last paper [1]. To broadly include data, only the word 'spondylolisthesis' was used for search, and updated till September 18, 2016. We selected original research data published after year 2000 and involving surgical treatment of DS, and reported from Japan (n=37 series, references 4-28), South Korea (n=11, references 29-38), mainland China (n=5 series, references 39-41), and ROC (Taiwan, n=3 series, references 42-43), America (n=20 series, references 44-57), and Europe (n=23 series, references 58-73). With the publications used for this analysis, European countries included Germany, United Kingdom, France, Italy, Spain, Sweden, Switzerland, and Norway. The mean ages of each patient series reported were extracted and used a single entry [Fig 1]. We tried our best to filter double/multiple reported data. It was noted that the series reported from mainland China often contain both congenital and degenerative spondylolisthesis. This lead to only five series could be used in this analysis.

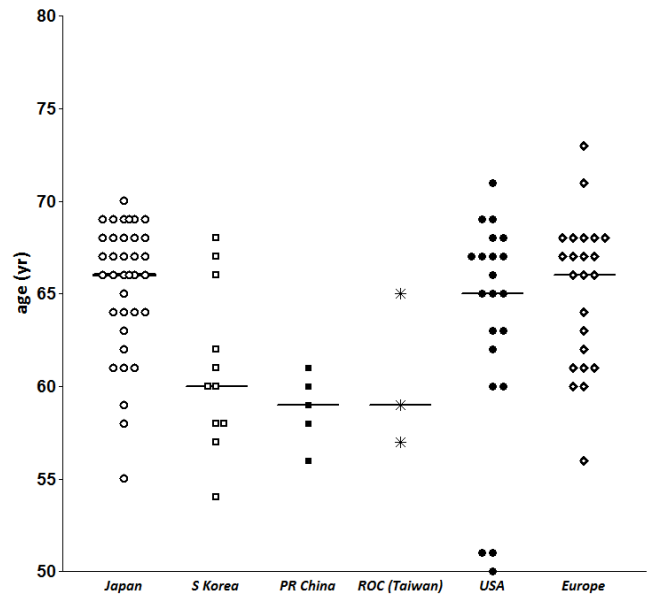

Fig 1. Degenerative spondylolisthesis patient age when undergone surgical treatment in six regions, i.e. Japan, South Korea, P.R. China, ROC (Taiwan), USA, and Europe. Each dot represents the mean age of one reported patient series.

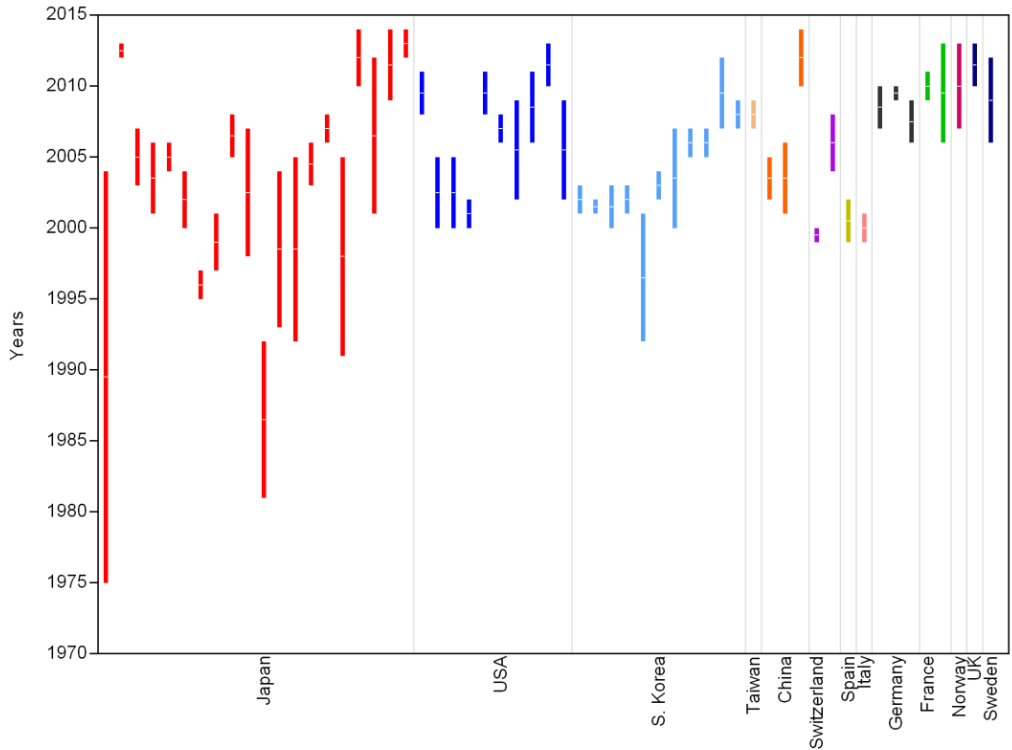

Fig 2. The surgery year of patient series contained in Figure 1 (extracted from references 4-73). Each bar represents one reported patient series.

The pooled results show median age of DS patients underwent surgical treatment was 66 years for Japan, 60 years for South Korea, 59 years for mainland China, 59 years for ROC (Taiwan), 65 years for USA, and 66 years for Europe (Figure 1). The median age of DS patients underwent surgical treatment in Japan, USA, and Europe was similar; while that of South Korea, mainland China, and maybe also (Taiwan), was younger. The surgery dates in the reports (Fig2, references 4-73) suggest the patient age differences between South Korea, Japan, USA, and Europe were unlikely caused by the difference of the years of operation.

This study represents a limited observation of published literatures. We did not perform statistical analysis, we also did not add weighting factor to each patient series according to study subject number. However, we believe the trend we saw in this study is likely to be real. This study indicates Japan and South Korea may have different surgical practice patterns. Our previous observation that Northeast Asians probably had surgical treatment earlier than Caucasians was likely not due to Northeast Asians manifest clinical symptoms earlier or more severe because of ethnic difference [1]. Instead, it was mainly due to DS patients in South Korea and China were more likely to undergo treatment at earlier age. It will be of interested to investigate whether more proportion of DS patients have been treated surgically in South Korea than in Japan. The cost-effectiveness of different approaches may require further analysis by professional spine surgeons.